# Multidisciplinary Cognitive Content Of Nanoscience and Nanotechnology

**Staša Milojević**

*Indiana University, School of Library and Information Science*

smilojev@indiana.edu

**Abstract** This article examines the cognitive evolution and disciplinary diversity of nanoscience/nanotechnology (nano research) as expressed through the terminology used in titles of nano journal articles. The analysis is based on the NanoBank bibliographic database of 287,106 nano articles published between 1981 and 2004. We perform multifaceted analyses of title words, focusing on 100 most frequent words or phrases (terms). Hierarchical clustering of title terms reveals three distinct time periods of cognitive development of nano research: formative (1981-1990), early (from 1991 through 1998), and current (after 1998). Early period is characterized by the introduction of thin film deposition techniques, while the current period is characterized by the increased focus on carbon nanotube and nanoparticle research. We introduce a method to identify *disciplinary components* of nanotechnology. It shows that the nano research is being carried out in a number of diverse *parent* disciplines. Currently only 5% of articles are published in dedicated nano-only journals. We find that some 85% of nano research today is multidisciplinary. The case study of the diffusion of several nano-specific terms (e.g., "carbon nanotube") shows that concepts spread from the initially few disciplinary components to the majority of them in a time span of around a decade. Hierarchical clustering of disciplinary components reveals that the cognitive content of current nanoscience can be divided into nine clusters. Some clusters account for a large fraction of nano research and are identified with such parent disciplines as the condensed matter and applied physics, materials science, and analytical chemistry. Other clusters represent much smaller parts of nano research, but are as cognitively distinct. In the decreasing order of size, these fields are: polymer science, biotechnology, general chemistry, surface science, and pharmacology. Cognitive content of research published in nano-only journals is closest to nano research published in condensed matter and applied physics journals.

## Introduction

Nanotechnology, alongside other converging technologies (such as biotechnology, information technology and cognitive sciences), is expected to have a tremendous impact on industry, society, human health, environment, sustainable development and security (Roco & Bainbridge, 2002b, 2005, 2001; Roco, Mirkin, & Hersam, 2011). Nanotechnology has variously been described as a "new frontier" (Barben, Fisher, Selin, & Guston, 2008), an "emergent field" (Wajcman, 2008), an "emergent, highly interdisciplinary field" (Zucker, Darby, Furner, Liu, & Ma, 2007), a "transdisciplinary research front" (Hayles, 2004) and a "rigorous scientific field" with "many signs of protodisciplinarity" (Milburn, 2004). It has formed at the intersection of the unprecedented number of disciplines, or fields, of science and engineering, blurring the lines between pure and applied research.

While there are similarities among all convergent technologies, such as the accelerated development, huge advances, multi-, inter- and cross-disciplinarity that affects traditional disciplines with which they are interacting (Roco, 2008), their specific character is different enough to make it difficult to compare them to each other directly. Toumey (2010) claims that the major impediment in comparing biotechnology and nanotechnology is their "grand commonality: both are broad, diverse families of technologies" (p. 475). Zucker & Darby (2005) compare the development of nanotechnology and biotechnology, using 1973 as a base year for biotechnology and 1986 as a base year for nanotechnology, and conclude that the development of instruments played crucial roles in the development of both of these fields.

Although nanotechnology is fairly young, its nature has already been intensely studied. Huang, Notten and Rasters (2011) examine more than 120 social science studies on nanotechnology, and find that 90% of those are based on the analysis of nanotechnology publications and patents, i.e., they use the bibliometric approach. Hullmann and Meyer (2003) provide an overview of such studies that attempt to "measure" nanotechnology. Focus of these studies includes: measuring interdisciplinarity of nanotechnology (Glänzel et al., 2003; Leydesdorff & Zhou, 2007; Lucio-Arias & Leydesdorff, 2007; Porter & Youtie, 2009; Rafols & Meyer, 2007; Schummer, 2004a, 2004b), studying collaboration patterns (Meyer & Bhattacharya, 2004; Milojević, 2010), exploring the relationship between nanotechnology and *nanoscience*[1] (Heinze, 2004; Meyer, 2001; Noyons et al., 2003), or the evaluation of the development of the field by country in which research is carried out (Huang, Chen, Chen, & Roco, 2004; Youtie, Shapira, & Porter, 2008). In addition, there have been a number of studies that have mapped nanoscience in order to identify thematic areas in the field (Bassecoulard, Lelu, & Zitt, 2007). And while some of the earliest studies of nanoscience focused on examining the growth of nanoscience/nanotechnology (Braun, Schubert, & Zsindely, 1997; Meyer & Persson, 1998), it is only recently that Li et al. (2008) provided a comprehensive longitudinal analysis of nanotechnology focusing on the period of its most rapid growth (1976-2004). Using a combination of methods (bibliographic analysis, content analysis and citation analysis) Li et al. analyzed the growth of the number of authors, level of contribution to nano research both by different countries and by individual institutions. In addition, Li et al. visualized major research topics and their evolution over time using the self-organization map algorithm.

Given the importance attached to interdisciplinarity for the advancement of science, it is not surprising that many of the above studies focused on measuring the interdisciplinarity of nanoscience. Empirical studies of interdisciplinarity of nanoscience used a variety of approaches to come to different conclusions. While some have argued that in nanoscience fields of research such as chemistry, physics, computer science, and biology merge strongly (Lucio-Arias & Leydesdorff, 2007), others have argued that the interdisciplinarity of nanoscale research is very weak. Namely, although nanoscale research encompasses many disciplines at equal rank, their research interaction is surprisingly low (Schummer, 2004a, 2004b). Schummer (2004b) also proposed that "a general measure of multidisciplinarity of a field is the number of disciplines involved" (p. 441). In the more recent study using science maps of nano articles and their references based on subject categories Porter and Youtie (2009) found that nanoscience research is highly integrative. Based on the review of the literature Huang, Notten & Rasters (2011) concluded that "nanotechnology is not a single homogeneous science or technology field, but a variety of nano-scale technologies spanning across various traditional disciplines" (p. 149).

One issue that these studies face is the lack of agreement on what interdisciplinarity, and related terms 'multidisciplinarity' and 'transdisciplinarity' mean and how exactly they differ from each other. In this study, we investigate nanotechnology's *multi*disciplinarity, rather than its *inter*disciplinarity. We define multidisciplinarity (adjective 'multidisciplinary') to describe research activities, problems, teachings, or bodies of knowledge (in other words the cognitive content) that has input from at least two *parent* scientific disciplines. According to this definition nanotechnology is a multidisciplinary field, with a number of parent disciplines. Parent disciplines of nanotechnology are various cognitively (or socially) delineated fields or disciplines in which nanotechnology research is being conducted (for example, applied physics, electrical engineering, physical chemistry, computer science). In none of them does nanotechnology represent the entire cognitive content of those parent fields. For example, only some part of research in applied physics will focus on nanotechnology, same for electrical engineering, etc. While we call these fields *parent* disciplines they can either reflect nano's true originating disciplines (e.g., materials science) or disciplines to which nano has "spread" later (e.g., biomedicine), but in either case are the result of nanotechnology's multidisciplinarity. The intersection of nanotechnology with parent disciplines will form what we call *disciplinary components* of nanotechnology. Therefore, for example, agriculture as a disciplinary component of nanotechnology will consist of the research on nanotechnology, but performed by researchers who work at agriculture departments or institutes or publish their work in journals with agriculture focus. As we will explain later, we determine disciplinary components by considering the primary topic of a journal in which nanotechnology research is published. This model of multidisciplinarity is shown in Figure 1. Note that the size of the circles is not meant to indicate the "size" of a parent field or of nanotechnology. Also, we consider parent fields as delineated, i.e., without overlaps, either between each other or within the disciplinary component. In other words, we do not consider the *inter*disciplinarity. Considering parent disciplines as separate entities is in no way meant to represent the actual inter-relationship between them, but defines how they will be treated in the context of this study. Finally, note that some fraction of a multidisciplinary field may not be covered by any of the parent fields. In the case of nanotechnology that would be the nano research that is published in journals with the sole nano focus.

---

[1] Terms nanotechnology and nanoscience are used interchangeably (often shortened to *nano*) to refer to a research field studying objects that have a size or structure of 1-100 nanometer, and we will use them in such a way in this paper as well.

Despite such a great interest in studying nanoscience its cognitive evolution and disciplinary diversity as expressed through terminology used in titles of nano journal articles has been neglected. Cognitive studies of science focus on science as a body of knowledge, that is, ideas and relationships between ideas. Given the importance of textual documents in the practice of science (Callon, Courtial, Turner, & Bauin, 1983; Latour & Woolgar, 1986), many scholars have focused on the shared conceptual systems of scientific communities as expressed through the *terminology* used in documents. Of the various components of textual documents, the *titles*, and the choice of words in them, are of particular importance. Leydesdorff (1989) claims that "word structure reflects internal intellectual organization in terms of the codification of word usage in the relevant disciplines" (p. 221). The analysis of words derived from document titles thus appears to be a promising approach to trace processes of discourse formation and cognitive structure of fields or disciplines, and is the approach we adopt for use in this study.

To summarize, the goal of this study is to determine the cognitive properties of nanoscience by studying the words and phrases that appear in nanoscience articles, and to trace how the cognitive content has changed since 1981. Furthermore, our study explores the disciplinary diversity (i.e., multidisciplinarity) of the cognitive content of nanoscience.

**Methods**

In this study we use several related methods of word analysis and apply each of them to a select representative group of title words and phrases from nanoscience research articles. Articles are selected to span over two decades (1981-2004), allowing us to study both the overall cognitive structure of the current nanoscience, and its evolution over this time period. We also study the differences in cognitive content with respect to disciplinary components, as well as with respect to different types of institutions with which nano-researchers are affiliated.

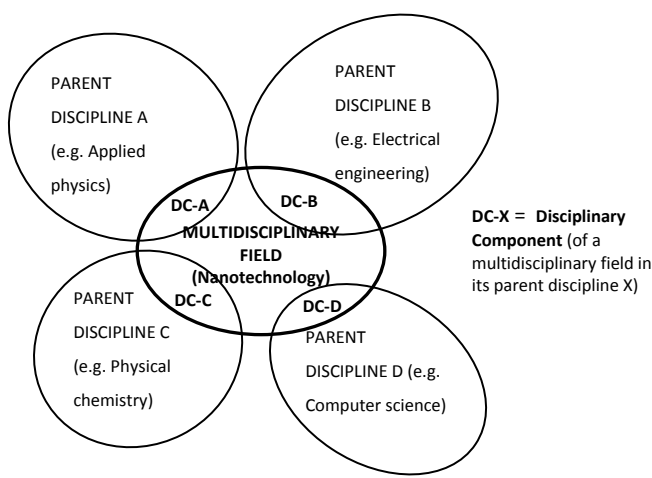

Figure 1. A model of multidisciplinarity used in this study.

**Data collection**

The principal source of data for this study is the article section of the NanoBank database (version Beta 1, released May 2007) (Zucker & Darby, 2007). NanoBank is a digital library of bibliographic data on articles, patents and grants in the field of nanotechnology (Zucker & Darby, 2005). Nanotechnology-related articles in NanoBank have been selected from the *Science Citation Index Expanded, Social Sciences Citation Index,* and *Arts and Humanities Citation Index* produced by Thomson Scientific (now Thomson Reuters). Two separate methods of selecting nano-related documents have been used in the creation of NanoBank (Zucker et al., 2007): (1) selecting articles that contain some of the 379 terms identified by subject specialists as being "nano-specific", and (2) selecting articles based on a probabilistic procedure for the automatic identification of terms. The database covers a 35-year period (1970-2004). While NanoBank contains entries dating back to the 1970s, before the nanoscience was recognized as a field, there are very few articles from this period and hence they cannot be analyzed reliably. Therefore, in this study we use data from year 1981 onward. This beginning point still predates the usually accepted emergence of nanotechnology in mid-1980s (Berube, 2006; Toumey, 2009), allowing us to explore its formative period.

We additionally filter NanoBank articles by requiring that the titles contain nano-specific keywords as identified by Porter et al. (2008). Specifically, we use keywords from their Table 2 listed as *MolEnv-I* and contained in bullets 1 through 8. NanoBank selected relevant articles by asking that a keyword appear either in title or in the abstract. However, abstracts are not available for all articles, especially prior to 1991. This can lead to unwanted biases and be problematic for the studies that compare different time periods. By performing this filtering on titles alone, we remove these effects. As a result, for the time range 1981-2004 we have 287,106 articles. Over this time period nanotechnology has grown exponentially, with the number of articles published annually increasing more than 20 times.

When analyzing trends, we will use the data from the entire time range (1981-2004). On the other hand, when discussing overall characteristics of nanotechnology we will focus only on the last five years of data (2000-04), i.e., what we refer to as the current period. Considering all articles (1981-2004), may appear as a better choice for characterizing the entire field, however, such strategy would actually fail to produce a balanced picture given the vast disproportionality in the number of articles at early and late times. Articles published in 2000-04 provide adequate characterization of the current state of the field, while accounting for 49% of all nano articles.

In order to obtain fuller understanding of the cognitive development of the field, we also study how the terminology was used by authors belonging to different institution types (e.g., university vs. industry). NanoBank, based on the contact information of the corresponding authors, provides the information on the type of the institution for most of the articles that we use in this study (97%). Table 1 provides the list of types of institutions as given by NanoBank.

Table 1 List of institution types in NanoBank.

|   | Code | Institution type |
|---|------|------------------|
| 0 | N/A  | Institution information not provided |
| 1 | FI   | Firm |
| 2 | UN   | University |
| 3 | NL   | National lab |
| 4 | RI   | Research institute |
| 5 | UG   | US government institute |
| 6 | HO   | Hospital |
| 7 | AS   | Academy of sciences |
| 8 | OT   | Other |

The overall distribution of the type of the institution for current nano articles (published between 2000 and 2004) is shown in Figure 2. The most dominant institution type is university (72%). Significantly behind are research institutes with 13%. Academies of sciences and firms contribute 6% and 4% respectively, while other categories are negligible.

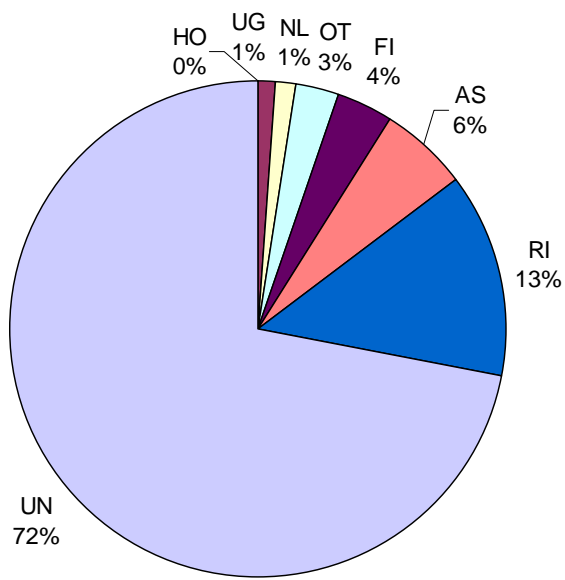

Figure 2. Distribution of articles by the type of institution of the corresponding authors (2000-04 period).

Data on parent disciplines in which nanotechnology research takes place are not part of the NanoBank database. We use subject categories of journals that publish nano-related research as a proxy for parent disciplines. To obtain this information we consult 2005 *ISI Journal Citation Reports* (JCR) to first obtain the subject categories of journals. A given journal can have one or more JCR subject category (total number of available subject categories is >200). We systematized various subject category combinations by identifying 42 parent disciplines in which nanotechnology research is published, and assigning one to each journal. Then, the parent discipline of any individual paper is simply determined by the parent discipline of the journal in which it was published. As a result of the application of this scheme each of 287,106 nano articles is assigned one of 42 parent disciplines, which therefore represent 42 disciplinary *components* of nanotechnology. Parent disciplines can more broadly be assigned to these areas: agriculture, biosciences, chemistry, computer science, engineering, earth sciences, ecology and environment, humanities, materials science, medicine, mathematics, nanotechnology[2], physics, physics & chemistry, multidisciplinary science and social sciences. Note that disciplinary components of nanotechnology are methodologically and conceptually different from nanotechnology's *research topics*, though there certainly exist parallels, as we will see later. In our case, we simply use the multidisciplinarity of nanotechnology to classify articles according to the field of the journal in which they appear.

Disciplinary components (and their codes), science areas to which they can be broadly grouped, as well as the number and the percentage of articles associated with each component (for 2000-04) are shown in Table 2.

Table 2. List of disciplinary components, broad areas to which they belong and corresponding number of articles (for 2000-04 period).

---

[2] As explained in the introduction, this refers to research that does not belong to any other parent discipline, i.e., to nano articles published in nano journals.

| No. | Code | Disciplinary component | Area | Articles | Art.(%) |
|---|---|---|---|---|---|
| 1 | A1 | Agriculture - all | AGRICULTURE | 171 | 0.1% |
| 2 | B1 | Biochemistry and molecular biology; cell biology; microbiology; biotechnology | BIOSCIENCES | 2844 | 2.0% |
| 3 | B2 | Biophysics | BIOSCIENCES | 1466 | 1.1% |
| 4 | B3 | Pharmacology | BIOSCIENCES | 1146 | 0.8% |
| 5 | B4 | Biosciences other (e.g. biology, zoology, plant science) | BIOSCIENCES | 196 | 0.1% |
| 6 | B5 | Biomedical engineering | BIOSCIENCES | 595 | 0.4% |
| 7 | C1 | Multidisciplinary chemistry | CHEMISTRY | 11157 | 8.0% |
| 8 | C2 | Analytical chemistry | CHEMISTRY | 11367 | 8.2% |
| 9 | C3 | Inorganic chemistry | CHEMISTRY | 1268 | 0.9% |
| 10 | C4 | Organic chemistry | CHEMISTRY | 955 | 0.7% |
| 11 | C5 | Electrochemistry | CHEMISTRY | 2151 | 1.5% |
| 12 | C6 | Chemistry - other | CHEMISTRY | 176 | 0.1% |
| 13 | CS1 | Computer science | COMPUTER SCIENCE | 546 | 0.4% |
| 14 | E1 | Electrical engineering | ENGINEERING | 3126 | 2.2% |
| 15 | E2 | Metallurgy | ENGINEERING | 2192 | 1.6% |
| 16 | E3 | Chemical engineering | ENGINEERING | 2081 | 1.5% |
| 17 | E4 | Engineering - other | ENGINEERING | 837 | 0.6% |
| 18 | ER1 | Earth sciences (geology, oceanography, meteorology…) | EARTH SCIENCES | 195 | 0.1% |
| 19 | EV1 | Ecology, environment, safety | ECOLOGY AND ENVIORNMENT | 658 | 0.5% |
| 20 | H1 | Humanities | HUMANITIES | 76 | 0.1% |
| 21 | M1 | Polymer science | MATERIALS SCIENCE | 8765 | 6.3% |
| 22 | M2 | Multidisciplinary materials science | MATERIALS SCIENCE | 12865 | 9.2% |
| 23 | M3 | Materials science - coatings & film | MATERIALS SCIENCE | 4559 | 3.3% |
| 24 | M4 | Materials science - ceramics | MATERIALS SCIENCE | 1992 | 1.4% |
| 25 | M5 | Materials science - other | MATERIALS SCIENCE | 500 | 0.4% |
| 26 | MD1 | Radiology | MEDICINE | 305 | 0.2% |
| 27 | MD2 | Medicine other | MEDICINE | 1054 | 0.8% |
| 28 | MT1 | Mathematics & statistics | MATHEMATICS | 845 | 0.6% |
| 29 | N1 | Nanoscience & nanotechnology | NANOTECHNOLOGY | 7493 | 5.4% |
| 30 | P1 | Multidisciplinary physics | PHYSICS | 9730 | 7.0% |
| 31 | P2 | Condensed matter & applied physics | PHYSICS | 22613 | 16.2% |

| 32 | P3 | Optics/microscopy | PHYSICS | 5527 | 4.0% |
| 33 | P4 | Crystallography | PHYSICS | 2382 | 1.7% |
| 34 | P5 | Mathematical physics | PHYSICS | 2766 | 2.0% |
| 35 | P6 | Physics of particles and fluids | PHYSICS | 1221 | 0.9% |
| 36 | P7 | Nuclear science/nuclear physics | PHYSICS | 1111 | 0.8% |
| 37 | P8 | Physics other (e.g. mechanics) | PHYSICS | 545 | 0.4% |
| 38 | PC1 | Physical chemistry; chemical physics; spectroscopy | PHYSICS & CHEMISTRY | 7577 | 5.4% |
| 39 | PC2 | Surface science | PHYSICS & CHEMISTRY | 1893 | 1.4% |
| 40 | PC3 | Physics and chemistry other (e.g. geochemistry & geophysics) | PHYSICS & CHEMISTRY | 226 | 0.2% |
| 41 | S1 | Science - multidisciplinary works | SCIENCE MULTIDISCIPLINARY | 1763 | 1.3% |
| 42 | SS1 | Social sciences | SOCIAL SCIENCES | 261 | 0.2% |
| 43 | XX1 | Unknown | UNKNOWN | 51 | 0.0% |

We see that most nano articles are published in the field of Condensed Matter & Applied Physics (P2), 16.2%. That is, the journals that publish most of the nano articles are classified as belonging to this field (specifically: *Physical Review B, Applied Physics Letters, Journal of Applied Physics,* etc.). This is followed by the fields of Multidisciplinary Materials Science (M2), 9.2%, Analytical Chemistry (C2), 8.2%, Multidisciplinary Chemistry (C1), 8%, and Polymer Science, (M1), 6.3%. Articles published in journals whose only or main focus is nanoscience (N1), such as *Physica E, Nano Letters, Nanotechnology* account for only 5.4% of all nano articles. Small percentage of articles published in nano-only journals demonstrates the highly multidisciplinary character of nanoscience, and warns against selection methods that select nanotechnology articles only from journals with primary nano focus. Such selection would be extremely incomplete, and as we will see later, rather unrepresentative. Finally, a small number of articles appear in journals classified as multidisciplinary science (S1). Again, this does not characterize the topics of these articles, but simply indicates the general profile of journals in which they are published. Journals *Science* and *Nature* belong to this category.

**Identification of the most frequently occurring nanoscience words or phrases**

We use word frequency to identify the most important *concepts* (and therefore, indirectly, *research topics*) in the field of nanotechnology. The main assumption underlying this approach is that "the most frequently appearing words reflect the greatest concerns" (Weber, 1990) (p. 51). Namely, the frequency with which a word appears is considered to be an indicator of "the importance of, attention to, or emphasis on" (Krippendorff, 2004) (p. 59) a particular word, or an idea or concept to which it is related.

Titles of NanoBank articles contain a total of 3,621,980 words (12.6 words per article title on average) of which 55,381 are unique (after excluding word variants). Word analysis software *WordStat* (Provalis Research, 2005) was used for lemmatization (consolidation of word variants, including identifying plurals and changing verb forms to infinitive), removal of numerals, and the removal of words belonging to a *stop list*, a list of 711 common English words have little value in cognitive analysis.

Certain words are most commonly found together with some other words, i.e., they form *phrases*. We would like to consider such phrases on par with individual words, but only if they are actually frequent enough to be ranked in the first 100.

There is no automated way of producing such mixed word/phrase list in *WordStat*, so we perform the following procedure. We first construct a list of most frequent individual words. Separately, using a built in procedure in *WordStat*, we identify most common *phrases* (combinations of individual words). We then insert the phrases occurring more frequently than the 100th ranked word to a temporary list of most frequently occurring words, but recalculating the frequencies of individual words and updating the ranking. We iterate until all phrase that can enter the top-100 list are inserted. This procedure resulted in the identification of 8 phrases for inclusion in the list of top 100 words and phrases. The list of terms[3], given in Table 3, provides the final list used in this study.

Table 3. List of the 100 most frequent words or phrases (bold) from article titles for the 1981-2004 time period.

| Rank | Term | Frequency |
|---|---|---|
| 1 | QUANTUM | 86507 |
| 2 | FILM | 24930 |
| 3 | EFFECT | 19532 |
| 4 | STRUCTURE | 19361 |
| 5 | SURFACE | 18870 |
| 6 | PROPERTY | 18273 |
| 7 | MOLECULAR | 17656 |
| 8 | POLYMER | 14476 |
| 9 | ELECTRON | 13447 |
| 10 | **THIN_FILM** | 13398 |
| 11 | NANOPARTICLE | 11816 |
| 12 | FIELD | 11574 |
| 13 | OPTICAL | 10791 |
| 14 | MAGNETIC | 10182 |
| 15 | SYSTEM | 10168 |
| 16 | SYNTHESIS | 10123 |
| 17 | GAAS | 9932 |
| 18 | SINGLE | 9879 |
| 19 | MATERIAL | 9608 |
| 20 | BASE | 9442 |
| 21 | **QUANTUM_DOT** | 9355 |
| 22 | LASE | 9321 |
| 23 | PHASE | 9275 |
| 24 | MODEL | 8914 |
| 25 | CHARACTERIZATION | 8852 |
| 26 | SI | 8609 |
| 27 | MOLECULE | 8342 |
| 28 | THEORY | 8187 |
| 29 | METAL | 7765 |
| 30 | MONOLAYER | 7731 |
| 31 | **CARBON_NANOTUBE** | 7620 |
| 32 | TEMPERATURE | 7444 |
| 33 | POLY | 7390 |
| 34 | GROWTH | 7386 |
| 35 | FULLERENE | 7234 |
| 36 | DYNAMIC | 6869 |
| 37 | ENERGY | 6729 |
| 38 | SILICON | 6603 |
| 39 | LAYER | 6503 |
| 40 | FORMATION | 6479 |
| 41 | NANOCRYSTALLINE | 6466 |
| 42 | SPECTROSCOPY | 6285 |
| 43 | OXIDE | 6260 |
| 44 | APPLICATION | 6162 |
| 45 | DIMENSIONAL | 6092 |
| 46 | ASSEMBLE | 6004 |
| 47 | TRANSITION | 6004 |
| 48 | COPOLYMER | 5983 |
| 49 | MECHANICAL | 5940 |
| 50 | ION | 5887 |
| 51 | METHOD | 5840 |
| 52 | SIMULATION | 5804 |
| 53 | INTERACTION | 5794 |
| 54 | SPIN | 5702 |
| 55 | INDUCE | 5666 |

---

[3] Hereafter we will refer to this combination of words and phrases as *terms*.

| | | |
|---|---|---|
| 56 | PROCESS | 5626 |
| 57 | DEPOSITION | 5622 |
| 58 | CARBON | 5592 |
| 59 | LOW | 5492 |
| 60 | BEAM | 5419 |
| 61 | CRYSTAL | 5366 |
| 62 | CHEMICAL | 5297 |
| 63 | LIQUID | 5239 |
| 64 | PREPARATION | 5230 |
| 65 | BIOSENSOR | 5168 |
| 66 | SEMICONDUCTOR | 5136 |
| 67 | TUNNEL | 5113 |
| 68 | NANOCOMPOSITE | 5103 |
| 69 | SOLUTION | 5056 |
| 70 | FE | 4992 |
| 71 | SCANNING | 4956 |
| 72 | ATOMIC_FORCE_MICROSCOPY | 4948 |
| 73 | PARTICLE | 4938 |
| 74 | GROW | 4933 |
| 75 | **LANGMUIR_BLODGETT** | 4899 |
| 76 | CU | 4886 |
| 77 | MECHANIC | 4786 |
| 78 | REACTION | 4742 |
| 79 | ELECTRONIC | 4731 |
| 80 | SUBSTRATUM | 4696 |
| 81 | WATER | 4633 |
| 82 | COMPLEX | 4620 |
| 83 | DOPE | 4560 |
| 84 | SCATTER | 4513 |
| 85 | INTERFACE | 4469 |
| 86 | ORGANIC | 4444 |
| 87 | CONTROL | 4437 |
| 88 | MEASUREMENT | 4422 |
| 89 | SIZE | 4412 |
| 90 | **MOLECULAR_DYNAMIC** | 4410 |
| 91 | MULTILAYER | 4407 |
| 92 | MICROSCOPY | 4369 |
| 93 | FORCE | 4357 |
| 94 | TRANSPORT | 4314 |
| 95 | ARRAY | 4286 |
| 96 | COUPLE | 4276 |
| 97 | ACID | 4248 |
| 98 | **SCANNING_TUNNEL_MICROSCOPY** | 4242 |
| 99 | SOLID | 4194 |
| 100 | **QUANTUM_CHEMICAL** | 4154 |

All analyses in this work are based on the 100 most frequent nanoscience terms. However, while 100 terms represent a small fraction of the total number of unique words appearing in titles, they still appear in the majority of the titles, and thus are representative of a bulk of the cognitive content of the field. This is demonstrated in Figure 3. Inclusion of the top 100 terms leads to representing 94% or the articles in the analysis. Since the curve is flattening, inclusion of more terms would not lead to any significant increases in the "coverage". Therefore, focusing on 100 terms is not only practical in terms of the analysis and the presentation of results, but ensures rather complete representation of the concepts present in article titles.

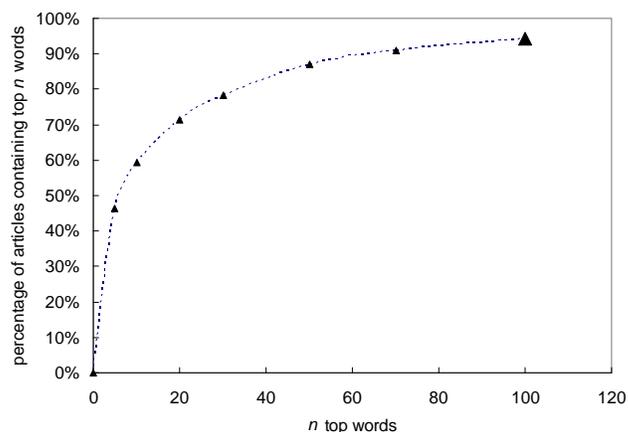

Figure 3. Fraction of article titles containing a given number of most frequent terms. Focusing on top 100 terms, as we do in this work, allows us to represent 94% of all article titles.

**Context of most frequent terms**

A look at the list of the most common terms (Table 3) points to the lack of context that terms by themselves can have. For example, the most frequent word QUANTUM, can appear in a number of contexts: (a) related to structures (e.g. inorganic semiconductor nanostructures: quantum wells, quantum wires and quantum dots[4]); (b) related to physical processes (e.g. the quantum Hall effect and the quantum confined Stark effect); and (c) related to applications (e.g. quantum cascade lasers). However, we can observe that in general the most frequent words refer to: structures, properties, materials, characterization and fabrication techniques, and devices. In Table 4 we list terms for which there is a context that occurs in more than 10% of cases of that term's occurrence. For example, the most common context for the term QUANTUM (after QUANTUM DOT and QUANTUM CHEMICAL which are treated as separate top 100 terms) is QUANTUM WELL (21.5%). Certain terms have more than one context in which they appear often. For example, the term FIELD, appears as MAGNETIC FIELD (18.9%), QUANTUM FIELD (16.5%), FIELD THEORY (15.7%), and ELECTRIC FIELD (10%). Similarly, the term BEAM, appears as MOLECULAR BEAM (55.4%), BEAM EPITAXY (46.7%), ION BEAM (15.6%), and ELECTRON BEAM (12.7%)

Table 4. Contexts of top 100 terms in which a context accounts for more than 10% of term occurrence.

| Term | Context | Percentage of context occurrence |
|---|---|---|
| QUANTUM | QUANTUM WELL | 21.5% |
| FILM | LANGMUIR BLODGETT FILM | 15.2% |
| STRUCTURE | WELL STRUCTURE | 12% |
| PROPERTY | OPTICAL PROPERTY | 13.1% |
| | MAGNETIC PROPERTY | 10.3% |
| MOLECULAR | MOLECULAR BEAM | 17% |
| ELECTRON | ELECTRON MICROSCOPY | 12.8% |
| FIELD | MAGNETIC FIELD | 18.9% |
| | QUANTUM FIELD | 16.5% |
| | FIELD THEORY | 15.7% |
| | ELECTRIC FIELD | 10% |
| OPTICAL | OPTICAL PROPERTY | 22.1% |
| MAGNETIC | MAGNETIC FIELD | 21.5% |
| | MAGNETIC PROPERTY | 18.5% |
| SYSTEM | QUANTUM SYSTEM | 13.9% |
| GAAS | GAAS QUANTUM | 15.9% |
| | GAAS ALGAAS | 13.6% |
| SINGLE | SINGLE WALL | 21% |
| | SINGLE QUANTUM | 13.4% |
| LASER | WELL LASER | 19.2% |
| PHASE | PHASE TRANSITION | 15.2% |
| MOLECULE | SINGLE MOLECULE | 11.4% |
| THEORY | FIELD THEORY | 22.2% |
| | QUANTUM THEORY | 20.2% |
| MONOLAYER | ASSEMBLE MONOLAYER | 31.6% |
| CARBON NANOTUBE | WALL CARBON NANOTUBE | 29% |
| TEMPERATURE | LOW TEMPERATURE | 24% |
| | ROOM TEMPERATURE | 16.2% |
| | HIGH TEMPERATURE | 11.7% |

---

[4] But note that we already treat QUANTUM DOT as a separate phrase.

| | | | | | | |
|---|---|---|---|---|---|---|
| POLY | POLY ETHYLENE | 18.9% | | | ION BEAM | 15.6% |
| DYNAMIC | QUANTUM DYNAMIC | 13.8% | | | ELECTRON BEAM | 12.7% |
| | | | CRYSTAL | LIQUID CRYSTAL | 21.4% |
| OXIDE | ETHYLENE OXIDE | 12.9% | | | SINGLE CRYSTAL | 18.1% |
| | OXIDE FILM | 11.6% | CHEMICAL | CHEMICAL VAPOR | 38.9% |
| DIMENSIONAL | TWO DIMENSIONAL | 45.4% | | | |
| | ONE DIMENSIONAL | 23% | LIQUID | LIQUID CRYSTAL | 21.8% |
| | | | | LIQUID CRYSTALLINE | 13.9% |
| | DIMENSIONAL QUANTUM | 17.9% | SEMICONDUCTOR | SEMICONDUCTOR QUANTUM | 19% |
| | THREE DIMENSIONAL | 17.2% | TUNNEL | SCANNING TUNNEL | 36.9% |
| ASSEMBLE | SELF ASSEMBLE | 93.4% | | TUNNEL MICROSCOPE | 31.1% |
| | ASSEMBLE MONOLAYER | 40.6% | | QUANTUM TUNNEL | 12.4% |
| TRANSITION | PHASE TRANSITION | 23.5% | SOLUTION | AQUEOUS SOLUTION | 20.7% |
| | TRANSITION METAL | 12.3% | SCANNING | SCANNING TUNNEL | 38.1% |
| COPOLYMER | BLOCK COPOLYMER | 31.5% | | SCANNING PROBE | 19.5% |
| | DIBLOCK COPOLYMER | 16.3% | | SCANNING ELECTRON | 11.4% |
| MECHANICAL | QUANTUM MECHANICAL | 55.4% | GROW | FILM GROW | 12.7% |
| | | | LANGMUIR BLODGETT | LANGMUIR BLODGETT FILM | 77.2% |
| | MECHANICAL PROPERTY | 18.5% | SOLID | SOLID STATE | 25.4% |
| ION | ION BEAM | 14.4% | MOLECULAR DYNAMIC | MOLECULAR DYNAMIC SIMULATION | 51.1% |
| SIMULATION | MOLECULAR DYNAMIC SIMULATION | 38.8% | | | |
| | CARLO SIMULATION | 13.3% | | | |
| SPIN | QUANTUM SPIN | 14% | | | |
| CARBON | CARBON FILM | 13.8% | | | |
| LOW | LOW TEMPERATURE | 32.6% | | | |
| | LOW ENERGY | 10% | | | |
| BEAM | MOLECULAR BEAM | 55.4% | | | |
| | BEAM EPITAXY | 46.7% | | | |

**Data analysis**

Two main forms of analyses are performed in this study: (a) relative word frequency and (b) multidimensional scaling analysis.

*Relative word frequency analysis*
Relative word frequencies are analyzed in order to describe the distribution of term occurrences across (a) disciplinary components, (b) institutions and (c) article publication years. Frequencies are relative because they do not depend on the total number of occurrences. They are calculated based on the percentage of a word's appearance in a given disciplinary component, institution, or time period. For example, if some term appears in 3% of titles in year A and the same percentage of titles in year B, it will have the same relative word frequency regardless of the absolute number of titles. We present relative word frequencies as heatmaps, in which different shades represent different relative frequencies (i.e., percentages). Heatmaps allow us to visualize trends in the usage of terms, and are produced using *WordStat*.

*Multidimensional scaling analysis*

We use multidimensional scaling technique to identify and visualize patterns in the underlying cognitive structure. Multidimensional scaling "is a generic name for a body of procedures and algorithms that start with an ordinal proximity matrix and generate configurations of points in one, two or three dimensions" (Jain & Dubes, 1988) (p. 46). Multidimensional scaling is a way of compressing inherent multidimensional properties of a dataset into fewer dimensions that can be visually displayed. Specifically, it makes it possible to visualize how certain objects cluster in terms of similarity, or how distant from each other (dissimilar) different clusters are. Each of the spatial dimensions resulting from scaling will correspond to some general concept present in the dataset, though it is not always possible to identify and describe these principal concepts. Multidimensional scaling also allows for comparison of relations between two different types of objects. For example, it will be used in this paper to explore the relations between terms and disciplinary components or between terms and publication years.

**Results**

**Evolution of the Cognitive Content of Nanoscience (1981-2004)**

*Relative frequency of top 100 terms*

To understand how the cognitive structure of nanoscience has changed over the formative period of the field, we analyze the temporal distribution of 100 terms. In Figure 4 we present the relative frequencies of term usage in different years using the heatmap.

Both the terms and the time periods have been hierarchically clustered using the similarity measure that groups items with similar relative frequencies together. Note that the hierarchical clustering preserved the original sequence of years. This implies that the evolution in the distribution of terms is a one-directional progression, with no instances of reversals to terms (and by extension to the corresponding topics or concepts) from previous time periods, even from year to year. Dendrogram formed by hierarchical clustering of years also allows us to see how closely related different time periods are. Those which are more similar will have a branching point closer to the year label, while those that differ the most will split closer to the "root" (top of the figure). From this dendrogram one can identify three distinct periods 1981-1990, 1991-1998, and 1999-2004, which branch close to the root, and are therefore more similar within each other than among each other. These periods should therefore correspond to three phases in the cognitive development of nanotechnology. Interestingly, the timings of these transitions coincide with two major National Science Foundation (NSF) initiatives: establishment of the first program for nanoparticle research "High Rate Synthesis of Nanoparticles" in 1991 and the first year of multidisciplinary initiative "Partnerships in Nanotechnology" in 1998 (NSF 1997).

Another way to explore the correlation between different periods is by using the multidimensional scaling. In Figure 5 we show 3D clustering of time periods with respect to the usage of terms. In this representation, similar groups will be clustered closer together in space. Also, the groups that are more like the overall sample will be closer to the origin. Figure 5 shows that the periods before and after 1991 are clustered apart from each other, meaning that they are qualitatively most different. This confirms what we observed in the dendrogram, but shows the extent of the discontinuity. After 1990 we also see that the sequence changes direction of progression, indicating a shift in topics studied. Since 1991 we see a progressive sequence with no gaps, signifying a more gradual change in term usage. A slight change in the direction of sequence again occurs after 1998 (corresponding to the third major branch in the dendrogram), but is not as drastic as one seen between 1990 and 1991.

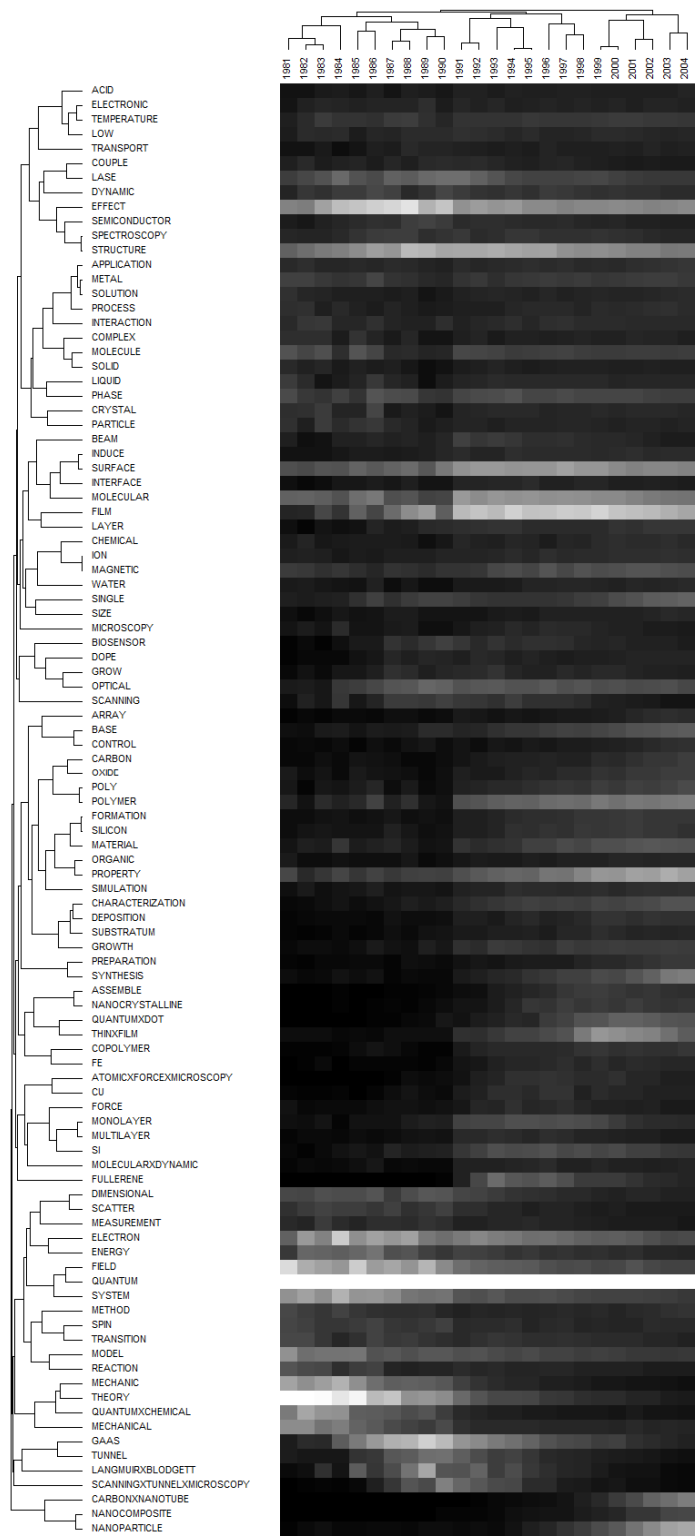

Figure 4. Heatmap of the relative frequencies of the 100 most frequently occurring terms across the time period from 1981 to 2004.

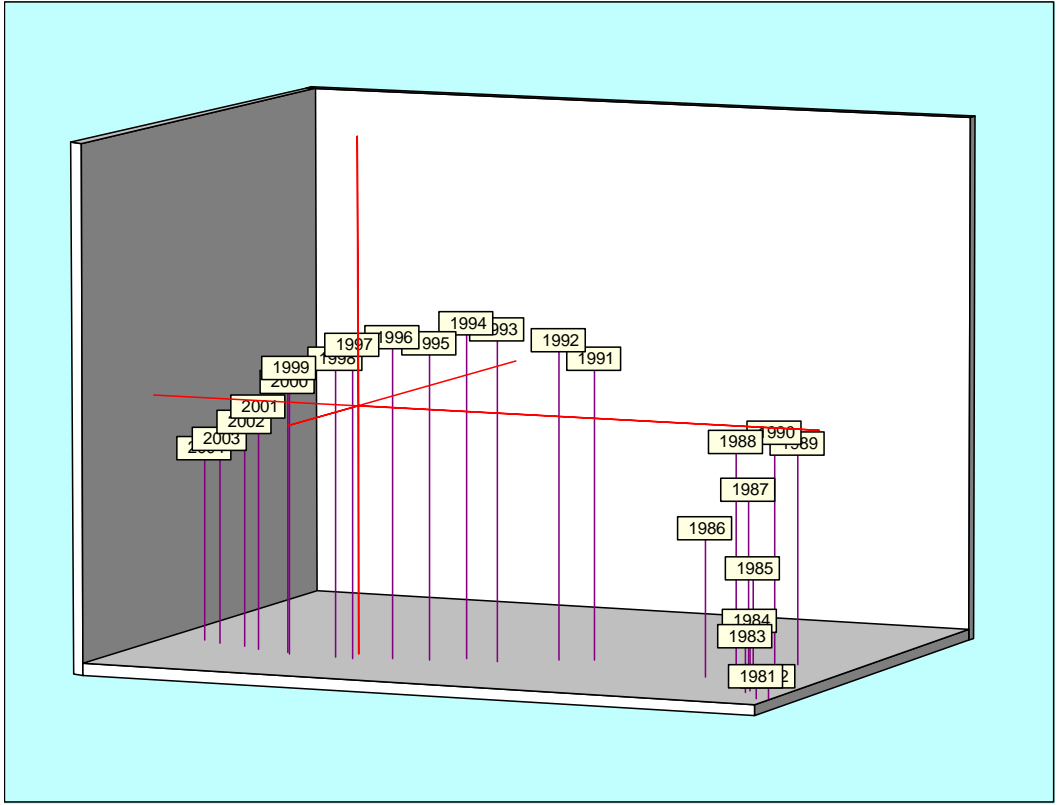

Figure 5. Spatial correlation of different time periods (years) based on the multidimensional scaling of 100 most frequently occurring article title terms.

Having established the distinct time periods of cognitive development of nanoscience, we now focus on how different terms have contributed to nanotechnology at different time periods (Figure 4). The only term that has had significant presence in all three time periods is QUANTUM. The terms that have been used more often in the early periods (1980s) and have experienced a drop in usage since then are: EFFECT, ELECTRON, ENERGY, FIELD, SYSTEM, THEORY, and QUANTUM MECHANICAL. Several terms have become dominant in the late 1980s and the early 1990s: STRUCTURE, GAAS, TUNNEL, LANGMUIR BLODGETT and SCANNING TUNNEL MICROSCOPY. The group of terms that started appearing in the 1990s and achieved their highest prominence then are: FILM, POLYMER, MATERIAL, QUANTUM DOT, PROPERTY, ASSEMBLE, THIN FILM, COPOLYMER, and FULLERENE. Finally, in the 2000s the words on the rise include CARBON NANOTUBE, NANOPARTICLES, NANOCOMPOSITE and SYNTHESIS.

*Diffusion of four nano-specific concepts*

The above analysis provides some insights into how certain terms have been used. To explore the process of a *diffusion*of a concept across different nanotechnology disciplinary components (or fields[5]), we will focus on four terms that represent key concepts in nanotechnology: CARBON NANOTUBE, THIN FILM, ATOMIC FORCE MICROSCOPE and SCAN(NING) TUNEL(LING) MICROSCOPY. The first two refer to nanotechnology materials and the second two to nanotechnology instrumentation.

**Carbon Nanotube**

Carbon nanotube is a member of the fullerene structural family. It is an allotrope of carbon with extraordinary strength, unique electrical properties and very good thermal conductivity. Although it is possible that the first carbon filaments were prepared in 1889 by Hughes and Chambers, it was due to the transmission electron microscopy that in 1952 Radushkevich and Lukyanovich showed the first evidence that carbon filaments are actually tubes, i.e. they are hollow. The interest in carbon nanofilaments/nanotubes has been present since then, but it was limited to the carbon materials science community. Much wider interest was aroused in 1991 "after the catalyst-free formation of nearly perfect concentric multi-wall carbon nanotubes … was reported as by-products of the formation of fullerenes via the electric-arc technique" (Monthioux et al., 2007) (p. 44). However, it was not until 1993 that the real breakthrough occurred with the discovery of single-wall carbon nanotubes. Given that the potential applications of nanotubes in electronics, optics and other fields of materials science seem countless and potentially revolutionary, it is not surprising that "about five papers a day are currently published by research teams from around the world with carbon nanotubes as the main topic" (Monthioux et al., 2007) (p. 44). Bearing in mind how active this research area is, it is not surprising that terms: CARBON NANOTUBE, (and its longer form [SINGLE and MULTI]-WALLED CARBON NANOTUBE) and CHEMICAL VAPOR DEPOSITION[6], are among the most frequently occurring terms in article titles in nanoscience/nanotechnology.

---

[5] In subsequent text we will occasionally shorten *disciplinary component* by *field*, but the meaning should always be clear from the context.

The term "carbon nanotube" (in singular and plural forms) occurs in 7,620 titles, appearing for the first time in several papers from 1992. For each year (starting with 1992) and each disciplinary component (Table 2) we find the percentage of articles containing this term. The results are shown as a heatmap in Figure 6 (white represents the maximum percentage in any cell, and corresponds to 16%). Disciplinary components are ordered according to the hierarchical clustering that will be discussed later on. The earlier time periods are at the bottom. The overall trend is for the term to diffuse (become more widespread) as time goes by. At the most recent time period (2004), the disciplinary components where this term is most present include Materials science – other (M5, 15.8%), followed by Multidisciplinary materials science (M2, 11.3%), and Nanotechnology (N1) with 10.7% of all titles. This, however, was not always the case. Prior to 2001, the disciplinary category in which the term CARBON NANOTUBE was present the most was general Science multidisciplinary works (S1). This probably reflects the fact that the discovery of carbon nanotubes was considered a large breakthrough and the pioneering researchers naturally sought wider audiences and greater acknowledgement by publishing in general science journals such as *Nature* or *Science*. Later on, this line of research became more mainstream, and gained its strongest foothold in certain specific disciplines. Also, while before 2000 the term was found only in several disciplinary components (e.g., Science multidisciplinary works (S1) and Multidisciplinary materials science (M2)), since then it has appeared in many other, and currently it contributes at levels above 1% in 30 fields and above 0.5% in 32 fields. The change in the number of disciplinary components with larger than 0.5% contribution is given in Figure 10. It shows the greatest increase in the last decade of the four terms analyzed here. Its absence is most notable in nano articles appearing in life sciences journals. Carbon nanotubes as a research area is certainly on the

**Thin Film**

Thin films are thin layers that are deposited onto surfaces of different materials, changing the properties of these materials. The underlying principles of various thin film deposition techniques have been "underrepresented in the literature, partly because much of thin-film technology has been developed empirically" (Smith, 1995) (p. xi). According to Smith (1995) major applications are in the areas of optics, electronics and pharmaceuticals, while the main disciplines involved in thin-film technology are: materials science, applied physics and electrical, mechanical and chemical engineering (Smith, 1995).

The term "thin film" is present in 13,398 article titles. The corresponding heatmap is shown in Figure 7 (white corresponds to 26%). The first article with the phrase "thin film" in title was published in 1973, although the context in which it appeared (wave physics) was not the one that thin films are currently associated with. The first rise.

---

[6] Chemical vapor deposition is one of the techniques that have been developed to produce larger quantities of nanotubes.

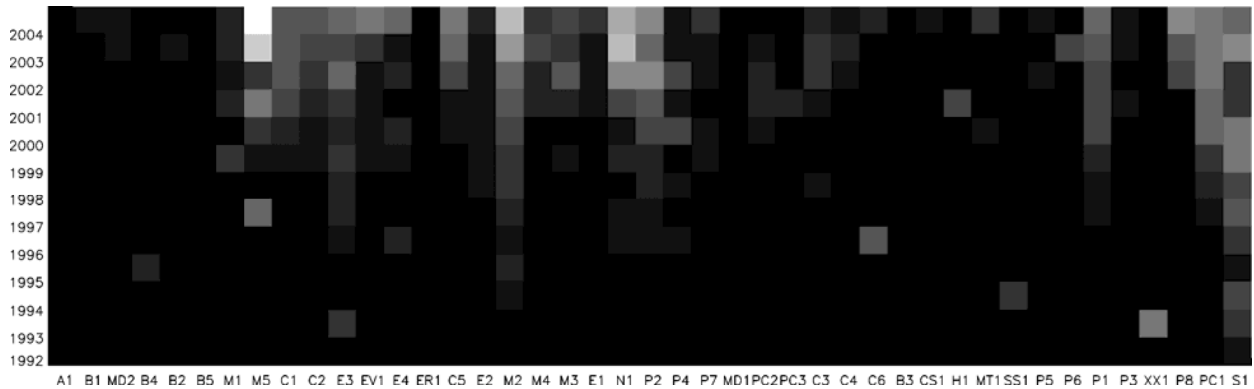

Figure 6. Heatmap of the phrase "carbon nanotube" as covered in different disciplinary components since 1992.

appearance in a modern context (in relation to materials) comes from 1978. Thin films apparently never made a splash in general science journals, or even nanotechnology journals. Their presence is more concentrated among different materials sciences fields and engineering (metallurgy and electrical engineering). So, for example in 1981 all articles in Metallurgy (E2) contained this term in their title. In a later period (in 1999 and 2001) ~25% of articles in Materials science – coatings & film (M3) and Electrical engineering (E1) contained this term in their titles. The number of disciplinary components where thin films accounted for more than 0.5% of the titles (Figure 10) has been rising steadily for most of the period of the study (in 2004 there are 28 such fields), though it appears to be leveling in recent periods. Currently, thin films are most often found in Materials science – coatings and film (M3), accounting for 17.9% of titles in this field, followed by Materials science – ceramics (M4) 13.3%, Surface science (PC2), 12.1% and Earth sciences (ER1), 11.6%.

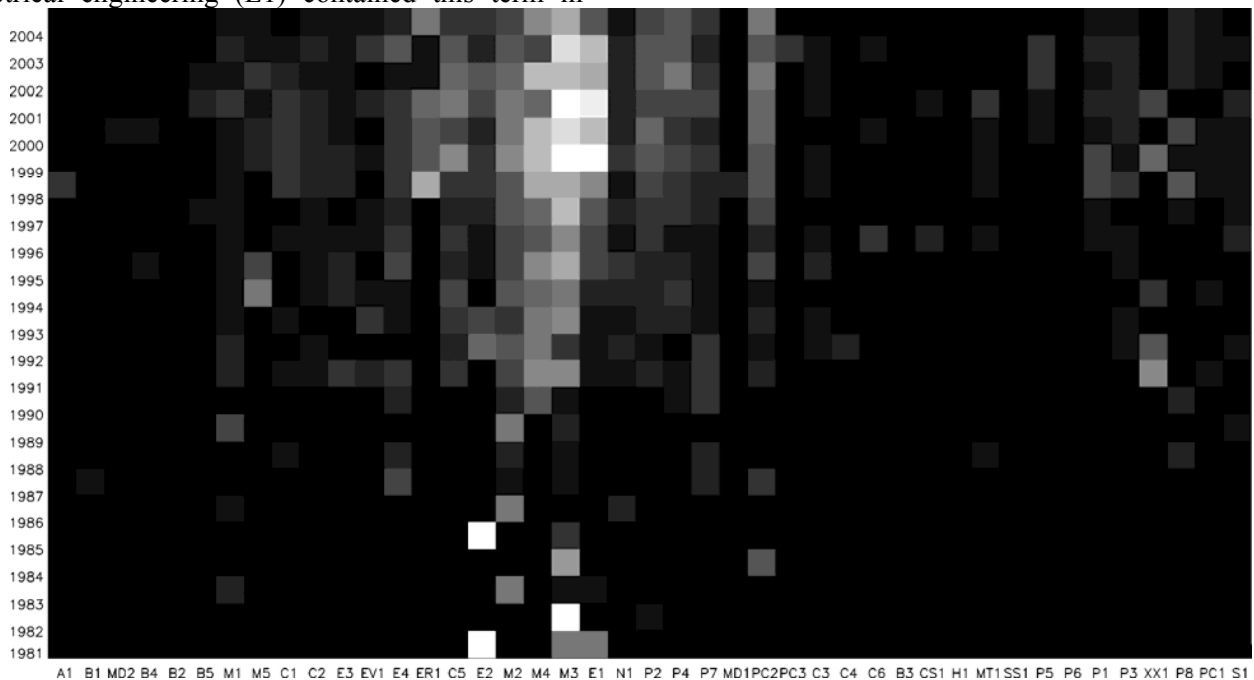

Figure 7. Heatmap of "thin film" as covered in different disciplinary components since 1981.

**Scanning Tunneling Microscopy**

A major invention that influenced the development of nanoscience and nanotechnology is the scanning tunneling microscope (STM). Jacoby (2000) observed:

> "Ask a dozen surface scientists to identify key developments in instrumentation that are responsible for catapulting nanotechnology to the front lines of physical science research. Nearly all of them will point to the advent of scanning probe microscopy." (p. 33)

STM was the first instrument "to enable scientists to obtain atomic-scale images and ultimately to manipulate individual atoms on the surfaces of materials" (Darby & Zucker, 2003) (p. 13). It was invented in 1979[7] by Gerd Karl Binning and Heinrich Rohrer at IBM Zurich "as a characterization tool in making thin films for a commercially important supercomputer" (Mody, 2006) (p. 60). Binning and Rohrer received the Nobel Prize in Physics in 1986 for their work on STM. In order to continue STM research after the original supercomputer project at IBM stopped, Binning and Rohrer, through alliances with academic researchers, worked on establishing STM replicators in different locations. The instrument was primarily used for research in the surface science. It was primarily used on metals and semiconductors, particularly silicon.

The term "scanning tunneling microscopy" and its variants (e.g., "microscope", "microscopic") along with its abbreviation (STM) are found in 9,005 article titles. The heatmap for this term is given in Figure 8 (white corresponds to 40%). The term first appeared within the field of Crystallography (P4) in 1982, and already had significant presence there (20% of titles). Also, given its early connection to IBM and supercomputers, it is not surprising that in 1986 81.8% nano-related articles from the field of Computer Science (CS1) had this term in their title. By 1990 STM spread to 29 disciplinary components (occurrence greater than 0.5%), but started to decline soon after (see also Figure 10). Its presence in Nanotechnology (N1) journals peaked in 1991 when 50% of the articles included the term in their title, but became much less prominent more recently, accounting for less then 1% of articles in the two latest time periods (2003 and 2004). During the peak period between 1988 and 1991, several fields (Electrochemistry C5, Physics and chemistry other PC3, Chemistry – other C6, Materials science – coatings & film M3, Surface science PC2, and Materials science – ceramics M4) had more than a quarter of titles mention the term. Nowadays, only 9 fields include this term in more than 1% of their titles and 18 fields include this term in more than 0.5% of titles, of which Surface science (PC2) is the most active area (15.8%).

The decline of the usage of the term (and obviously the instrument) is closely tied to the invention of a more encompassing instrument, the atomic force microscope (ATM), in 1986 and its adaptation for the use on living cells in 1991.

**Atomic Force Microscopy**

The STM's limitation to applications on metal and semiconductors had an effect on the extent of its usage. Gerd Binning (IBM Zurich) and Calvin Quante (Stanford University) were interested in carving out "interdisciplinary niches for STM" (Mody, 2006) (p. 64). During Binning's sabbatical year at Stanford, 1985-1986, Quante, Christopher Gerber and he invented a new, enhanced instrument, the atomic force microscope (AFM). The AFM "broadened the range of materials which could be viewed at the atomic scale and enhanced the ability to manipulate individual atoms and molecules" (Darby & Zucker, 2003) (p. 13-14). The AFM has been modified in 1991 for use on living cells.

The term "atomic force microscopy", its variants (e.g., microscope, microscopic) and abbreviated form AFM is encountered in 8,581 titles. Its heatmap is given in Figure 9 (white corresponds to 24%). The term is first mentioned in a paper from 1986, matching the discovery year. In the next 10 years the term gained prominence in a large number of nano disciplinary components, though some fields have accepted it sooner than others. The term was steadily mentioned in general science journals, but at a relatively modest rate. Nanotechnology journals briefly paid increased attention to the term (1994-96), but then the interest died down. The field of Physics and chemistry other (PC3) was the first nano disciplinary component to have had significant presence of the term in its article titles, especially in the period between 1989 and 1998 (reaching the peak in 1991 with 50% of its titles having the term). The field of Medicine other (MD2) had the highest presence of the term in its titles between 1994 and 1998 (between 14 and 24%). Although the prevalence of the term has somewhat diminished since that time, it is still very strong in the most recent periods. Finally, the field with the highest presence of the term in the most recent time period, Biophysics (B2), has witnessed frequent usage of the term in its titles since 1996 (between 15% and 23% of its titles having the term). Currently, it is "bioscience" fields that are most interested in the term, Biophysics (B2), Earth sciences (ER1), Biochemistry and molecular biology (B1) and Agriculture (A1). After spreading very quickly to 28 fields (at levels of 0.5% or higher) in 1993, the number of fields has stayed relatively constant, between 30 and 37 (Figure 10).

---

[7] Mody (2006) reports that it was invented in 1979; Jacoby (2000) places the invention of the STM in the early 1980s; while Darby (2003) say that it was invented in 1981.

Bearing in mind that the term represents an instrument, it is safe to say that the AFM has been more widely used than its predecessor STM, especially since 1999. Given its capability to manipulate and view live cells it is not surprising that the term is so prominent among the researchers in the field of biophysics.

From Figure 10 we can see how quickly and to what extend have the four terms have diffused across disciplinary components (fields). The rate of spread of both STM and ATM was similar, and was faster than for the other two terms. On average, terms take about a decade to reach maximum spread, at which time they are found in some 3/4 of the disciplinary components.

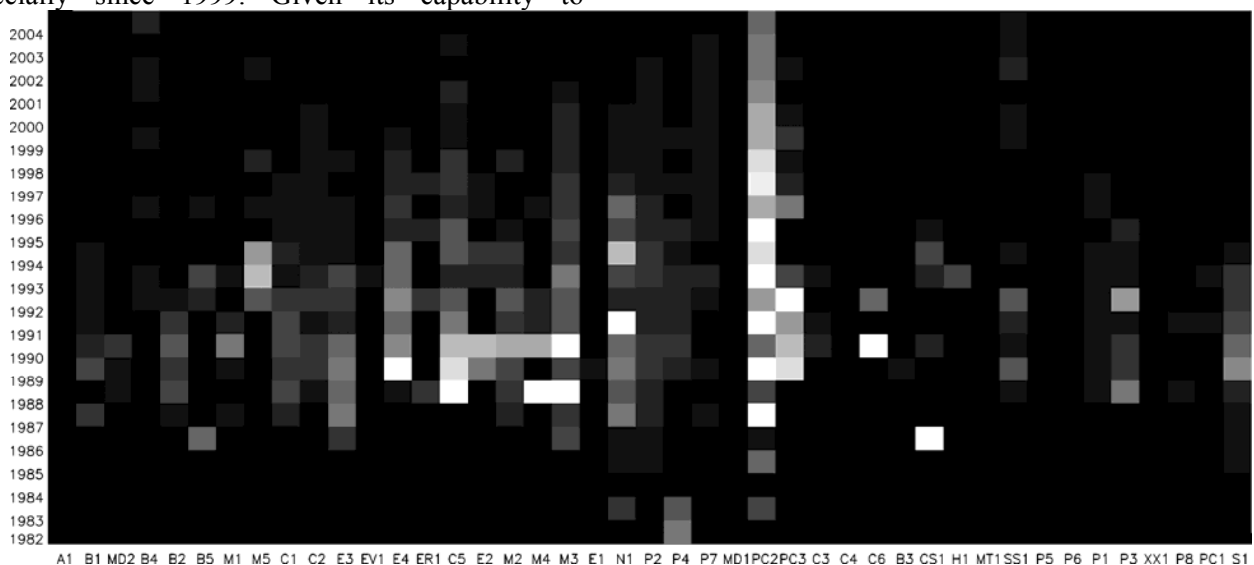

Figure 8. Heatmap of "scanning tunneling microscopy" as covered in different disciplinary components since 1982.

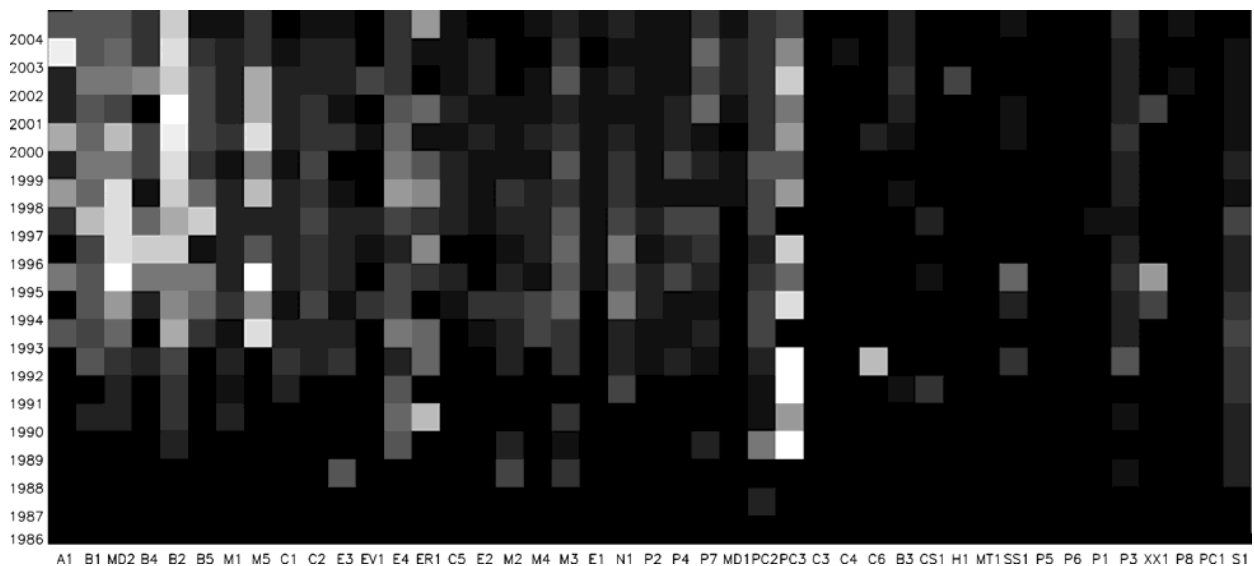

Figure 9. Heatmap of "atomic force microscopy" as covered in different disciplinary components since 1986.

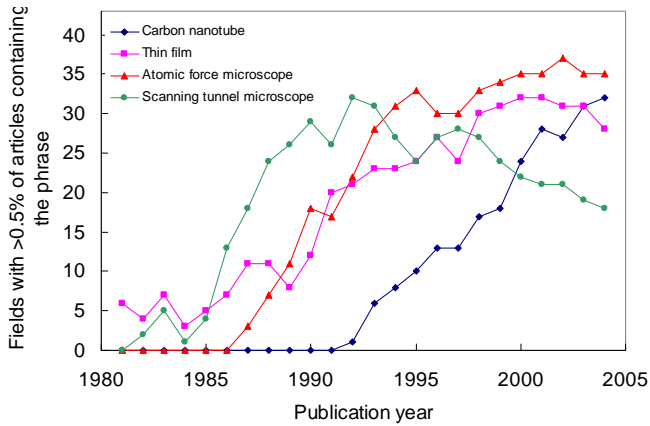

Figure 10. The change in the number of disciplinary components (fields) with phrases "carbon nanotubes", "thin film", "atomic force microscopy" and "scanning tunneling microscopy" being found in more than 0.5% of the titles.

**Differences in cognitive content with respect to nanoscience disciplinary components – the cognitive structure of nanoscience**

Given that nanoscience/nanotechnology articles are published in journals that belong to a large number of parent disciplines (what we call nano disciplinary components), we next turn our attention to the distribution of the most frequent title terms across those disciplinary components. Since we are no longer interested in the dynamics, the analysis focuses on the current period (2000-04). For each disciplinary component we obtain relative frequencies of the 100 top terms and show it as a heatmap (Figure 11, grayscale saturates at 2%). Again, both the rows (representing terms) and the columns (disciplinary components) have been hierarchically clustered so that those that have similar distributions appear closer together. Heatmap reveals several concentrations (bright regions) where groups of words contribute significantly to a number of fields. Such examples include LASER and OPTICAL being present in Optics/microscopy (P3) and Electrical engineering (E1); CU, FE (abbreviations for metals Copper and Iron), and NANOCRYSTALLINE being prominent in Metallurgy (E2), or the terms ATOMIC FORCE MICROSCOPY, MOLECULAR, and BIOSENSOR related to Agriculture (A1), all Bioscience fields (B1-B5), and Medicine (MD2).

Next we focus on the branching of disciplinary components based on the relative frequency of top 100 terms, i.e., on interpreting the dendrogram at the top of the heatmap in Figure 11. The branching points in the dendrogram of disciplinary components are difficult to read in Figure 11, so in Figure 12 we reproduce the dendrogram at larger scale. We can use the branching in the dendrogram to divide the fields in some number of clusters. The idea behind this is that the clusters will tell us about related disciplinary components, and therefore help us determine the cognitive *structure* of nanotechnology. The dendrogram structure suggests grouping the disciplinary components into 9 clusters (separated by short dashed lines in Figure 12). The disciplinary components comprising these groups, the total number (and percentage) of articles that each of these groups contributes to the dataset, and the list of ten most common terms in each group together with the relative frequency of their occurrence in that group are shown in Table 5.

*Multidisciplinary physics and fringe fields.* Starting from the bottom of the dendrogram, we focus on the first cluster, which contains some of the apparently most closely "related" fields (i.e. their branching points are very close to value of the similarity index of 1; the left side of the dendrogram): Mathematics (MT1), Social Sciences (SS1), Humanities (H1), and Computer Science (CS1). While some of these fields represent very different *parent* disciplines, the reason they all appear similar is that they are all "fringe" fields in relation to the rest of the nanotechnology disciplinary components, which apparently makes them use similar terms in their titles. Apart from these fringe fields that account for a small percentage of articles, the cluster is dominated by (in decreasing order of number of articles) Multidisciplinary Physics (P1), Physical chemistry/Chemical physics/Spectroscopy (PC1), Optics/Microscopy (P3), Mathematical Physics (P5) and Science Multidisciplinary (S1). This is the second largest cluster accounting for 22% of all the nano-papers. By far the dominant term in this group is QUANTUM (51% of all article titles in this cluster contain this term), and it is this term that connects fringe fields to various physics disciplines. It is followed by terms EFFECT (5.5%), FIELD (5.4%), MOLECULAR (5.4%), SYSTEMS (5.0%).

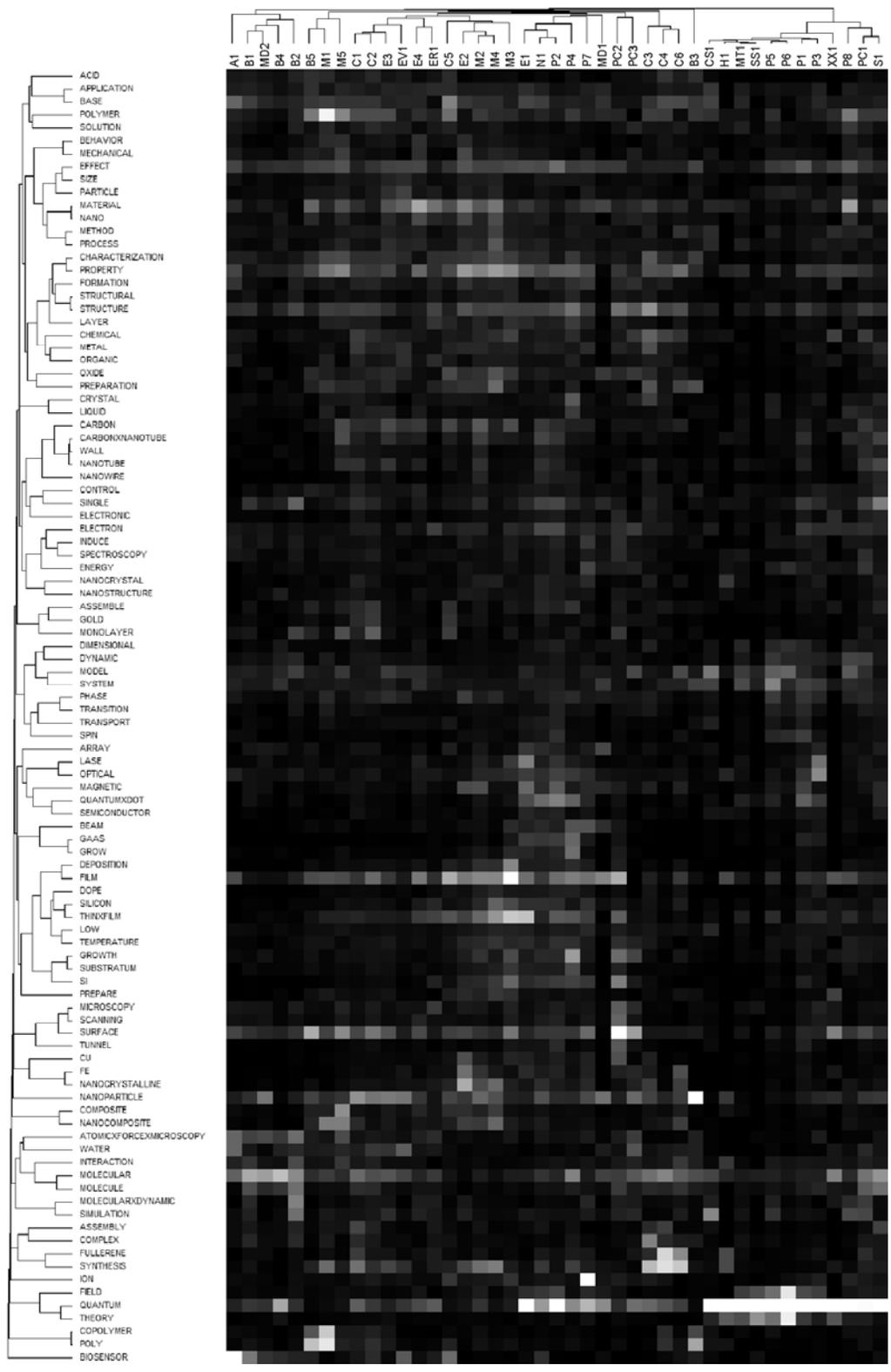

Figure 11. Heatmap of 100 top words and nanoscience disciplinary components.

*Pharamcology*. The second and the smallest branch consists of only one field, Pharmacology (B3), and accounts for only 0.8% of all the nano-papers. The most common term in this group is NANOPARTICLE (39%), followed by POLY (18%), DRUG (17%), DELIVERY (16%), and LIPID (12%). Thus, despite its small size, this cluster has very distinct terms usage that connects nanotechnology with classical pharmacology.

*General chemistry*. The third branch consists of three chemistry-related fields (in decreasing size): Inorganic Chemistry (C3), Organic Chemistry (C4) and Chemistry Other (C6). This cluster accounts for 1.7% of nano articles, making it one of the three smallest clusters. These fields often use terms SYNTHESIS (22%), MOLECULAR (13%), FULLERENE (13%), STRUCTURE (12%) and QUANTUM (10%).

*Surface science*. The fourth branch consists of two fields of which Surface Science (PC2) dominates and the other is Physics and Chemistry other (PC3). Together they account for 1.5% of articles, making it the second smallest branch. It is characterized by frequent usage of terms SURFACE (30%), FILM (16%), SI (11%), STRUCTURE (11%), GROWTH (11%), and STM (10%).

*Condensed matter and applied physics, plus core nano*. The fifth branch is the largest (27% of all nano articles) and consists of two dominant disciplinary components: Condensed Matter and Applied Physics (P2) and Nanoscience (N1). The first accounts for 61% of all articles in this cluster, while the latter makes 20%. Smaller fields in this cluster include Electrical Engineering (E1), and Crystallography (P4). These fields most often use the term QUANTUM (22%), followed by QUANTUM DOT (10.7%), FILM (9.5%), EFFECT (8.9%), PROPERTY (8.3%), and THIN FILM (8.1%). Articles published in journals classified primarily as nanotechnology are most closely related to the disciplinary component of Condensed matter & applied physics (P2). This is not surprising given that this disciplinary component clearly dominates in terms of the fraction of all papers, and thus represents *core nano parent discipline*. This also shows that the profile of journals that are classified as mostly "nanotechnology", and therefore core nano research, is closer to physics than it is to chemistry.

*Materials science*. The sixth branch consists of fields: Multidisciplinary Materials Science (M2), Materials Science –Coatings & Film (M3), Metallurgy (E2), Electrochemistry (C5), and Materials Science – Ceramics (M4). It is the fourth largest cluster (17% of articles). The fields in this group often use terms: FILM (17%), PROPERTY (13%), THIN FILM (12%), SYNTHESIS (8.1%), MATERIAL (8.0%), and NANOPARTICLE (7.6%).

*Analytical chemistry*. The seventh and third largest branch consists of fields that include: Analytical Chemistry (C2), Multidisciplinary Chemistry (C1) and Chemical Engineering (E3) accounting for 19% of all articles. These fields most often use terms NANOPARTICLE (13%), SURFACE (9.2%), MOLECULAR (8.8%), FILM (7.8%), POLYMER (7.4%), and SYNTHESIS (7.4%).

*Polymer science*. The eighth branch consists of three fields: Polymer Science (M1), Biomedical Engineering (B5), and Materials Science other (M5), where the latter two have a smaller contribution to the cluster. All three account for 7.1% of articles. These fields most often use terms: POLYMER (24%), POLY (23%), COPOLYMER (20%), NANOCOMPOSITE (13%), PROPERTY (12%), SYNTHESIS (11%) and POLYMERIZATION (10%).

*Biotechnology*. The final branch consists of five fields, with major contributions from: Biochemistry and Molecular Biology (B1), Biophysics (B2) and Medicine other (MD2), and some smaller contribution from Agriculture (A1). Altogether this cluster makes 4.1% of all articles. This cluster is characterized by frequent usage of the term MOLECULAR (17%), followed by PROTEIN (11%), BIOSENSOR (9.2%), DNA (9.0%), CELL (8.3%), and ATOMIC FORCE MICROSCOPY (8.2%). The term BIOSENSOR is particularly dominant in Agriculture.

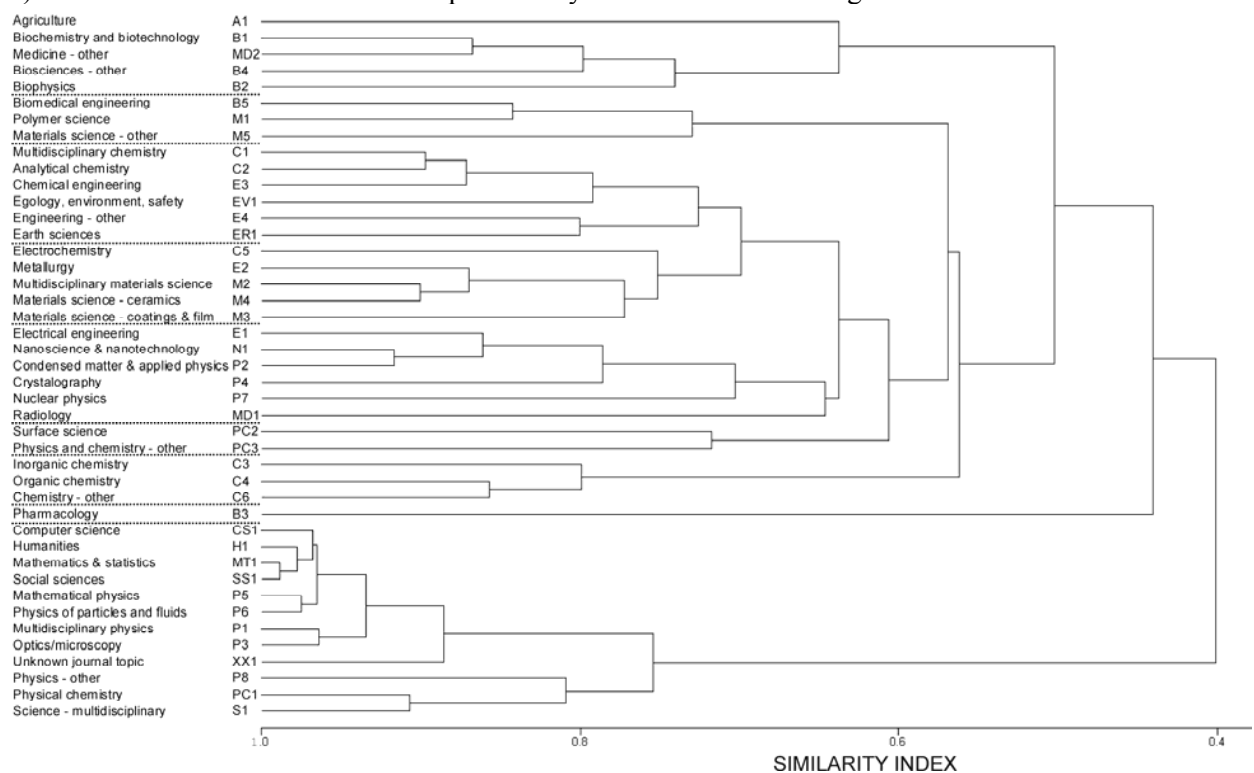

Figure 12. Dendrogram of nanoscience disciplinary components based on title words similarity index.

Table 5. List of disciplinary clusters of nanoscience determined from hierarchical clustering of its disciplinary components. First column gives the name of the cluster (bold corresponds to those making more than 10% of nano articles). Second column gives the percentage of articles and size rank of the cluster. Third column lists disciplinary components in a given cluster and contribution of each in percents. Fourth column gives a list of ten most frequent terms in each cluster and the percentage of articles belonging to that cluster in which the term is found (2000-04 period).

| **PHYSICAL CHEMISTRY & FRINGE FIELDS** | 22.2% (2) | Multidisciplinary physics 31.5%<br>Physical chemistry 24.5%<br>Optics/microscopy 17.9%<br>Mathematical physics 8.9%<br>Science - multidisciplinary 5.7%<br>Physics of particles and fluids 4.0%<br>Mathematics & statistics 2.7%<br>Computer science 1.8%<br>Physics - other 1.8%<br>Social sciences 0.8%<br>Humanities 0.2% | 1 QUANTUM 50.8%<br>2 EFFECT 5.5%<br>3 FIELD 5.4%<br>4 MOLECULAR 5.4%<br>5 SYSTEM 5.0%<br>6 THEORY 4.7%<br>7 SINGLE 4.6%<br>8 OPTICAL 4.4%<br>9 STRUCTURE 4.4%<br>10 MOLECULE 4.4% |
| PHARMACOLOGY | 0.8% (9) | Pharmacology 100% | 1 NANOPARTICLE 38.7%<br>2 POLY 17.9% |
| | | | 3 DRUG 16.9%<br>4 DELIVERY 16.3%<br>5 LIPID 11.5%<br>6 SOLID 10.2%<br>7 PREPARATION 8.4%<br>8 MOLECULAR 8.4%<br>9 LOAD 8.3%<br>10 COPOLYMER 7.7% |
| GENERAL CHEMISTRY | 1.7% (7) | Inorganic chemistry 52.9%<br>Organic chemistry 39.8%<br>Chemistry - other 7.3% | 1 SYNTHESIS 22.2%<br>2 MOLECULAR 13.1%<br>3 FULLERENE 12.7%<br>4 STRUCTURE 12.3%<br>5 QUANTUM 10.2%<br>6 COMPLEX 9.8%<br>7 CHEMICAL 9.1%<br>8 PROPERTY 9.0%<br>9 CHARACTERIZATION 7.3%<br>10 BASE 6.8% |
| SURFACE SCIENCE | 1.5% (8) | Surface science 89.3%<br>Physics and chemistry - other 10.7% | 1 SURFACE 29.9%<br>2 FILM 15.9%<br>3 SI 11.4%<br>4 STRUCTURE 11.1%<br>5 GROWTH 10.6%<br>6 STM 10.1%<br>7 SCANNING 9.5%<br>8 MICROSCOPY 8.9%<br>9 THIN FILM 8.8%<br>10 TUNNEL 8.4% |
| **APPLIED PHYSICS & CORE NANO** | 26.6% (1) | Condensed matter & applied physics 61.1%<br>Nanoscience & nanotechnology 20.2%<br>Electrical engineering 8.4%<br>Crystalography 6.4%<br>Nuclear physics 3.0%<br>Radiology 0.8% | 1 QUANTUM 22.0%<br>2 QUANTUM DOT 10.7%<br>3 FILM 9.5%<br>4 EFFECT 8.9%<br>5 PROPERTY 8.3%<br>6 THIN FILM 8.1%<br>7 MAGNETIC 7.2%<br>8 STRUCTURE 7.1%<br>9 ELECTRON 6.2%<br>10 SINGLE 6.0% |
| **MATERIALS** | 17.1% | Multidisciplinary materials science | 1 FILM 16.9% |
| **SCIENCE** | (4) | 54.1%<br>Materials science - coatings & film 19.2%<br>Metallurgy 9.2%<br>Electrochemistry 9.1%<br>Materials science - ceramics 8.4% | 2 PROPERTY 12.7%<br>3 THIN FILM 11.8%<br>4 SYNTHESIS 8.1%<br>5 MATERIAL 8.0%<br>6 NANOPARTICLE 7.6%<br>7 CARBON 7.3%<br>8 NANOCRYSTALLINE 7.2%<br>9 EFFECT 6.6%<br>10 SURFACE 6.4% |
| **ANALYTICAL CHEMISTRY** | 18.9%<br>(3) | Analytical chemistry 43.2%<br>Multidisciplinary chemistry 42.4%<br>Chemical engineering 7.9%<br>Engineering - other 3.2%<br>Egology, environment, safety 2.5%<br>Earth sciences 0.7% | 1 NANOPARTICLE 12.6%<br>2 SURFACE 9.2%<br>3 MOLECULAR 8.8%<br>4 FILM 7.8%<br>5 POLYMER 7.4%<br>6 SYNTHESIS 7.4%<br>7 QUANTUM 7.0%<br>8 STRUCTURE 6.0%<br>9 PROPERTY 6.0%<br>10 MONOLAYER 5.9% |
| POLYMER SCIENCE | 7.1%<br>(5) | Polymer science 88.9%<br>Biomedical engineering 6.0%<br>Materials science - other 5.1% | 1 POLYMER 24.4%<br>2 POLY 23.2%<br>3 COPOLYMER 19.6%<br>4 NANOCOMPOSITE 12.6%<br>5 PROPERTY 12.1%<br>6 SYNTHESIS 10.5%<br>7 POLYMERIZATION 10.4%<br>8 BLOCK 8.4%<br>9 FILM 8.4%<br>10 SURFACE 8.0% |
| BIOTECHNOLOGY | 4.1%<br>(6) | Biochemistry and biotechnology 49.6%<br>Biophysics 25.6%<br>Medicine - other 18.4%<br>Biosciences - other 3.4%<br>Agriculture 3.0% | 1 MOLECULAR 16.5%<br>2 PROTEIN 11.1%<br>3 BIOSENSOR 9.2%<br>4 DNA 9.0%<br>5 CELL 8.3%<br>6 AFM 8.2%<br>7 MOLECULE 7.9%<br>8 SINGLE 5.9%<br>9 QUANTUM 5.7%<br>10 NANOPARTICLE 5.6% |

Finally, we can refer to the dendrogram (Figure 12) to determine how distant other clusters are from the *core nano* cluster by following the sequence of branching points. Closest to core nano are *materials science* and *analytical chemistry* clusters. These are followed by *surface science, polymer science* and *general chemistry* clusters, in that order. *Biotechnology* and *pharmacology* clusters are still further away from core nano. Finally, the cluster containing various multidisciplinary physics fields and many fringe disciplines sets itself apart the most. Interestingly, this most distant cluster accounts for one fifth of nano titles, which may warrant further investigation.

**Terms and the Type of Institution**

Most nanoscience/nanotechnology research is done at universities, followed by research institutes, academies of science and firms (Figure 2). In order to examine the intellectual structure of nanoscience/nanotechnology as a function of the type of institution at which research is done, we study the distribution of the most frequent title terms within each institution type.

Institution type is a property of an author and not necessarily of a paper (which can be coauthored by authors from different types of institutions), but we will here suppose that the institution type of the corresponding author will dominate the cognitive content in the article title. Figure 13 shows the heatmap of 100 most frequent terms and nine institution types (full names of institution types are given in Table 1). Terms and institution types are again ordered according to the similarity in distributions. Most terms are distributed equally in various institution types. This is not surprising given that the disciplinary components themselves contribute at the similar level in all of the institution types except hospitals (Milojević, 2009). We find that hospitals are the most dissimilar with respect to other institutions, so they branch off at an early stage (Figure 14). This difference comes both from the lack of presence of the term QUANTUM (which is by far the most dominant term in all other institutions) and the significant presence of the term MOLECULAR in research done in hospitals. Somewhat more distinct from others are the titles of articles with no organization recorded. Interestingly, the terms QUANTUM and NANOPARTICLE have significant presence here, indicating that papers with institution type missing are perhaps not just a random grouping of articles. Other institution types are relatively similar in their use of terms, with universities and research institutes being the most closely related.

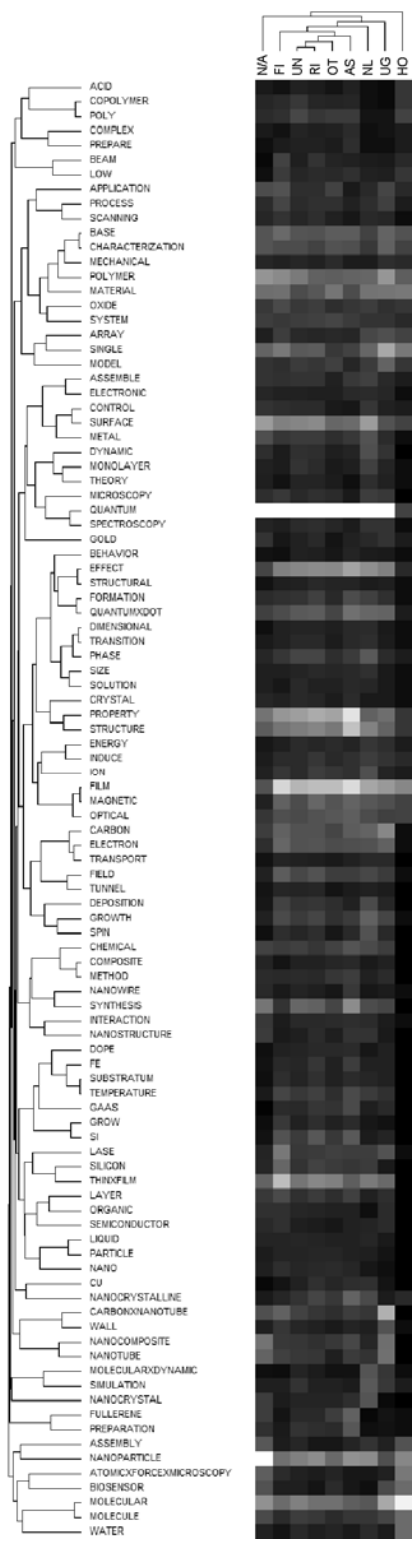

Figure 13. Heatmap of 100 top terms and institution types.

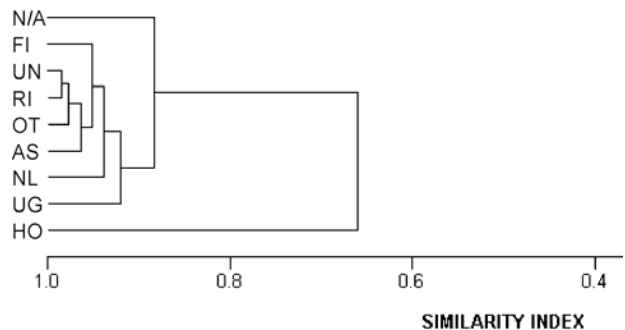

Figure 14. Dendrogram of institution types based on title terms similarity index.

**Summary and conclusions**

This study provides a description of the cognitive content of the field of nanoscience/nanotechnology from the perspective of frequently occurring title words or phrases, follows the change of that content over a period of more than two decades, and explores the multidisciplinary character of nanotechnology and through it its cognitive structure.

During the period that we covered in this study (1981–2004), nanoscience research experienced an exponential growth, with 20 times as many nano-related articles being published at the end of this period as at the beginning. The findings show that the evolution in the distribution of terms in nanoscience has presented a continual progression in multidimensional scaling space, with no instances of reversal to terms (and by extension to the topics or concepts) from the previous epochs. This result is expected from such young and rapidly progressing field as nanotechnology, so its confirmation in our analysis also serves as an indicator of the soundness and the robustness of the methods used.

Furthermore, both the hierarchical clustering of publication years with respect to the trends of usage of title terms, and the above-mentioned multidimensional scaling analysis indicate three distinct time periods of cognitive development of nanoscience: 1981-1990, 1991-1998 and 1998-2004. Between the first two periods there is a marked jump in patterns of term usage, while the second transition is milder, and leads more to the change in direction. The three periods can be identified with the following epochs in the development of nanoscience: formative epoch, establishment epoch and the current epoch.

The 1980s period can be considered the beginnings of nanoscience/nanotechnology as a field. This period can be characterized as the formative stage of the field, during which the key conceptual and technological advances were achieved in order to allow later development. The terminology used in this period was still not standardized and was different from the one used in nanoscience today. To strengthen the argument that this is the period in which the field was setting up not only its human capital but its knowledge base, the literature on the history of the field indicates that it is during this period that the major scientific breakthroughs in nanotechnology happened: the major instrument Scanning Tunneling Microscope (STM) was invented in the early 1980s, only to be enhanced and to a certain extent replaced by another instrument Atomic Force Microscope (AFM) in 1986. Another major discovery, that of "Buckminster fullerenes", was made in 1985. Many of the thin film deposition techniques were being developed in this period as well. The prefix "nano" appeared in article titles from 1980.

The nanoscience/nanotechnology started coalescing around common terminology only around 1991, the same time it became structurally cohesive in the *social* realm, by forming large connected networks of researchers (Milojević, 2009). Based on this we can claim that only after 1990 we can start talking about nanoscience as a scientific field, and our analysis indicates that it is then that the second, establishment phase begins. A look at the literature on the history of nanoscience/nanotechnology indicates that the second time period (1991-1998) also coincides with the period in which a number of nanoscience institutions were formed. The first nanoscience-related journals, *Nanotechnology* and *Nanostructured Materials*, were established in 1990 and 1992 respectively. The first nanoscience/nanotechnology-related conferences started being held during this time period. For example, the Foresight Institute has sponsored annual international conferences on nanotechnology since 1989. The for-profit Nano Science and Technology Institute was formed in 1992. But perhaps the most significant development that ushered this phase was that the NSF started its first initiative on nanoparticle research, "High Rate Synthesis of Nanoparticles", in 1991.

After the establishment epoch our analysis indicates another shift around year 1999, which may represent the start of the mature phase of nanoscience. This period is marked by the increased focus on concepts related to carbon nanotubes and nanoparticles. The beginning of this period follows right after the first year of NSF's multidisciplinary initiative "Partnership in Nanotechnology" from 1998. At the same time, the international community of nano researchers came up with the new, conceptually different, definition of nanotechnology (Roco, 2011). Also, we note that the first doctoral program in nanotechnology was offered in 2002, a feature of an already well-established field.

Our data are complete only to year 2004. It would be interesting to see whether nano has since experienced another cognitive shift that would signify a fourth phase, and if so, when did that shift happen. First, regarding the overall development, there is an indication that nano has entered a more stable phase because the growth of nano publications has slowed down since 2005 (Porter & Youtie, 2009). Using different methods but similar data, Porter & Youtie (2009) have identified: physical chemistry, multidisciplinary materials science, multidisciplinary chemistry, applied physics, and condensed matter physics as the main research areas within nano in 2008. Since these are the same areas that we identify as important prior to 2004, their results suggest that there has not been a major shift in focus since 2004. However, Roco (2011) recently projected a major shift in nano research in 2011-2020, and noted a significant and rapid increase in the publications on "active nanostructures" since 2005. Definitive answers will require that we apply the same methodology used for our current data to more recent data. We are planning to address this in a future work

Nanoscience is widely accepted to have a multidisciplinary character (research being done within many independently established fields), and is likely also interdisciplinary (various fields interacting among each other). Both have proved challenging to explore quantitatively. In this work we focus on nanoscience's multidisciplinarity, by using the fact that nano research is being published in a variety of journals whose main topic is not nano research. This allows us to identify how nanoscience is divided into various disciplinary components that themselves are also a part of some other parent discipline (where a parent discipline can be either some of the originating fields or a field to which nano spread afterwards). Using this method we identify 42 disciplinary components of vastly different sizes, the most dominant of which is the condensed matter and applied physics. While nanotechnology appears firmly established as a field, the above measure indicates that it is still largely multidisciplinary, with 95% of nano research from the recent period being published in journals whose focus is not only on nanoscience. This number may be somewhat artificially boosted by a possible social phenomenon. Namely, nano researchers may be prone to continue publishing in journals with a broader focus, as a consequence of the earlier epochs when nano did not have its dedicated venues for publishing. This tendency may be intentional as well, as it may still be important to present nano work to a broader audience. Our clustering analysis shows that the vocabulary of article titles work published in nano-only journals is most similar to that of papers published in condensed matter and applied physics journals. So the two components together can be considered to be core nano research. Since these account for 22% of all nano articles, it follows that the rest is multidisciplinary. In either case, the multidisciplinarity remains high, around 85%.

The wide variety of journals publishing nano-related research and lack of dominance of journals solely devoted to nanotechnology is one indicator of field's multidisciplinarity. Another indicator of its multidisciplinarity is the existence of fairly distinctive research groups, each coalescing around different set of topics as expressed by article title words. The clusters themselves to great extent mirror the disciplinary boundaries of parent fields.

The largest cluster that we identify is the one that contains articles appearing in nanotechnology-only journals. Actually, the disciplinary component of nanotechnology is the closest in terms of terminology used to condensed matter & applied physics (which is the field with the highest production of nanotechnology articles) followed by electrical engineering. We consider this to be a *core nano* cluster and it accounts for one quarter of all nano papers. Closest in cognitive respect to this core nano cluster are two clusters both relatively large in size: that of *materials science* and of *analytical chemistry*. The first also contains disciplines such as metallurgy and electrochemistry, and the other also contains chemical engineering. These two together make more than one third of all nano articles. Next in relatedness to core nano cluster is a small cluster (2% of total articles) containing articles from the fields related to *surface science*, followed by *polymer science* and *general chemistry* clusters together accounting for one tenth of the nano corpus. Less related to core nano is the small cluster of *biotechnology* fields (4% of articles), which includes biochemistry & molecular biology, biophysics and medicine. Nanoscience research in *pharmacology* cluster is still further removed from core nano, but despite being the smallest cluster (1%), it is very well defined in focus.

The cluster that is furthest apart according to the patterns of terms usage contains the remaining one fifth of the articles. In it we find research from "fringe" disciplines in relation to core nanoscience: e.g., Mathematics, Social Sciences, Humanities, Computer Science, and also a number of physics disciplines that apparently do not carry out the research considered to be core for nanotechnology. Interestingly, despite its fringe character, this is the second largest cluster. It is mostly connected through the very high usage of a single term *quantum*.

It is safe to conclude that there are indicators that different terminology is dominant in different disciplinary components, with clear clustering among physics (which includes core nano), chemistry, material science, and biosciences.

Finally, we have also shown that the nanoscience research reported in the journals mostly comes from scientists affiliated with universities, but the way in which they use nano concepts does not differ much from other types of institutions, such as the industry. The only exceptions are hospitals, where the focus is on the molecular concepts.

The methodology applied in this work is a combination of bibliometric and cognitive content analysis methods applied to the most frequent title words and phrases, and as such represents a novel way to study a scientific field. The results of such an analysis allows us to construct a robust, objective picture, which can then be compared with more subjective accounts on the development of the field and its cognitive structure. Due to the limitations of the input database our study of nanoscience ends with year 2004. Future studies will be able to apply our methods to subsequent periods and observe any further cognitive evolution in this field.

Acknowledgements: Certain data included herein are derived from NanoBank (Lynne G. Zucker and Michael R. Darby, NanoBank Data Description, release 1.0 (beta-test), Los Angeles, CA: UCLA